\documentclass[12pt]{article}
\usepackage{amsmath}
\usepackage[dvips]{graphics}

\begin{document}

\title{The ``physical process version" of the first law and the
generalized second law for charged and rotating black holes}

\author{Sijie Gao and  Robert M. Wald \\Enrico Fermi Institute and
Department of Physics \\   
University of Chicago \\
5640 S. Ellis Avenue \\
Chicago, Illinois 60637-1433}
\maketitle

\begin{abstract}
We investigate both the ``physical process'' version of the first law
and the second law of black hole thermodynamics for charged and
rotating black holes. We begin by deriving general formulas for the
first order variation in ADM mass and angular momentum for linear
perturbations off a stationary, electrovac background in terms of the
perturbed non-electromagnetic stress-energy, $\delta T_{ab}$, and the
perturbed charge current density, $\delta j^a$.  Using these formulas,
we prove the ``physical process version" of the first law for charged,
stationary black holes. We then investigate the generalized second law
of thermodynamics (GSL) for charged, stationary black holes for
processes in which a box containing charged matter is lowered toward
the black hole and then released (at which point the box and its
contents fall into the black hole and/or thermalize with the ``thermal
atmosphere'' surrounding the black hole). Assuming that the thermal
atmosphere admits a local, thermodynamic description with respect to
observers following orbits of the horizon Killing field, and assuming
that the combined black hole/thermal atmosphere system is in a state
of maximum entropy at fixed mass, angular momentum, and charge, we
show that the total generalized entropy cannot decrease during the
lowering process or in the ``release process''. Consequently, the GSL
always holds in such processes. No entropy bounds on matter are
assumed to hold in any of our arguments.

\end{abstract}


\section{Introduction}   \label{introduction}

The close mathematical and physical connection between the laws of
black hole physics and the laws of thermodynamics provides the main
foundation for ideas and speculations on the nature of quantum gravity
in the strong field regime. Many aspects of black hole thermodynamics
are on a completely firm foundation, such as the classical laws
of black hole mechanics and the fact that black holes radiate via the
Hawking process as perfect black bodies (of finite size) at
temperature 
\begin{eqnarray} 
T_H = \frac{\kappa}{2 \pi}
\label{TH}
\end{eqnarray} 
where $\kappa$ denotes the surface
gravity of the black hole (see, e.g., \cite{tbh} for a recent
review). Nevertheless, there remain some unresolved and/or
controversial issues in black hole thermodynamics.

One relatively minor unresolved issue concerns the ``physical process
version'' of the first law of classical black hole mechanics for
charged black holes. Consider a linear perturbation of a
stationary, electrovac black hole corresponding to taking one to
another stationary, electrovac black hole. Then, as originally shown
by Bardeen, Carter, and Hawking \cite{bch} (see \cite{iyer1} for a
generalized version) the first order variations of the area $A$, mass
$M$, angular momentum $J$, and charge $Q$ are related by 
\begin{eqnarray} 
\frac{1}{8 \pi} \kappa \delta A = \delta M-\Omega_H \delta J-\Phi_{\rm bh} \delta Q
\label{pverf} 
\end{eqnarray} 
where $\Omega_H$ denotes the angular velocity of the horizon and
$\Phi_{\rm bh}$ denotes the electrostatic potential of the horizon
(i.e., $\Phi_{\rm bh} = - A_a \xi^a$, where $\xi^a$ is the horizon
Killing field). However, it also is possible to consider a ``physical
process'' wherein some charged matter is thrown into an initially
stationary, electrovac black hole. Assuming that the black hole
eventually settles down to a final stationary state, one may calculate
the change in black hole area, $\delta A$, using the Raychaudhuri
equation and compare it with $\delta M$, $\delta J$, and $\delta
Q$. If eq.(\ref{pverf}) were to fail, this would give rise to an
inconsistency with the assumption that the black hole settles down to
a final stationary state, and would thereby provide strong evidence
against cosmic censorship. Conversely, a proof of the ``physical
process'' version of the first law would provide support for cosmic
censorship.

A proof of the ``physical process'' version of the first law for
uncharged black holes was given in \cite{waldbook}. However, some
difficulties arise in extending this proof to the charged
case\footnote{We are indebted to A. Ashtekar for pointing out these
difficulties to us.}. One of the purposes of this paper is to remedy
these difficulties by showing that eq.(\ref{pverf}) holds for all
physical processes.

A crucial issue in black hole thermodynamics is the validity of the
generalized second law (GSL), which states that the total generalized
entropy $S^\prime \equiv S + S_{\rm bh}$ never decreases \cite{bek},
where $S$ is the ordinary entropy of matter outside the black hole
and, in general relativity, $S_{\rm bh} = \frac{1}{4}A$. Early arguments
by Bekenstein for the validity of this law in quasi-static lowering
processes required the assumption that ordinary matter must obey
an entropy bound of the form \cite{bek2} 
\begin{eqnarray} S\leq 2\pi ER
\label{entb} 
\end{eqnarray} 
in order to prevent the box from being lowered too close to the black
hole. An alternative resolution not requiring any entropy bounds on
matter was given by Unruh and Wald \cite{unruhwald}, taking into
account the quantum buoyancy force of the thermal atmosphere
surrounding the black hole. This analysis has been criticized by
Bekenstein on a variety of grounds \cite{bek3}-\cite{bek5}; see
\cite{uw2} and \cite{mark} for responses to \cite{bek3} and
\cite{bek4}. Recently, it has been argued that even stronger entropy
bounds than (\ref{entb}) are needed for charged and rotating black
holes \cite{bm}-\cite{hod2}. These arguments have been countered for
charged black holes by Shimomura and Mukohyama \cite{muko}.

In view of the above situation, it seems worthwhile to give a new,
general analysis of the validity of the GSL in quasi-static lowering
processes that is applicable to charged and rotating black holes and
invokes no model dependent assumptions concerning the thermal
atmosphere and the contents of the box, or assumptions about the size
and shape of the box (other than that it is much smaller than the
black hole but large enough that a thermodynamic treatment of the
thermal atmosphere is adequate).  In this paper we shall give such an
analysis. Our key assumptions are as follows:
\begin{enumerate}
\item The thermal atmosphere admits a suitable local thermodynamic
description with respect to observers following orbits of the horizon
Killing field, $\xi^a$. Furthermore, the thermal atmosphere is in
thermal equilibrium with itself. More precisely, we cannot increase
the entropy of the thermal atmosphere by any rearrangement of it that
keeps fixed its total mass, angular momentum, and charge, as well as
other conserved quantities, such as the number of particles of a given
species.
\item The thermal atmosphere is in thermal equilibrium with the black
hole at temperature eq.(\ref{TH}). More precisely, we cannot increase
the total generalized entropy of the black hole/thermal atmosphere
system by any rearrangement that keeps fixed the total mass, angular
momentum, and charge of the total system.
\end{enumerate}

We consider processes in which a box containing arbitrary matter and
charge is quasi-statically lowered toward the black hole and then
``released'', so that the box or its contents are dropped into the
black hole and/or allowed to thermalize with the thermal
atmosphere. We will show that if the contents of the box are in
thermal equilibrium, no decrease in the total generalized entropy can
occur during the ``lowering phase'' of a quasi-static process.
However, in the ``release phase'', the total mass, angular momentum,
and charge do not change. Consequently, since the black hole/thermal
atmosphere system is assumed to have maximum generalized entropy at
fixed mass, angular momentum, and charge, the generalized entropy
cannot decrease in the ``release phase'' either.

A key ingredient in our analysis of both the ``physical process''
version of the first law and the GSL is a general formula for the
variation of ADM mass, $\delta M$, and angular momentum, $\delta J$,
for perturbations of stationary or, respectively, axisymmetric
electrovac spacetimes. In section 2, we will prove that these
quantities are given by\footnote{Note that for perturbations of
Minkowski spacetime (with $\Sigma$ taken to be a slice so that
$\partial \Sigma$ is empty), these equations reduce to the frequently
used---but seldom, if ever, derived!---formulas $\delta M =
-\int_\Sigma\epsilon_{dabc} t^e \delta {T^d}_e$ and $\delta J=
\int_\Sigma \epsilon_{dabc} \varphi^e \delta {T^d}_e$.}
\begin{eqnarray} 
\delta M
=-\int_\Sigma\epsilon_{dabc}\left(t^e \delta {T^d}_e + A_e t^e\delta
j^d \right) + \int_{\partial \Sigma} \left( \delta{\mbox{\boldmath $Q$}} [t]-t \cdot
{\mbox{\boldmath $\Theta$}} \right)
\label{dmem}
\end{eqnarray}
\begin{eqnarray}
\delta J= \int_\Sigma \epsilon_{dabc}\left(\varphi^e \delta {T^d}_e +
A_e \varphi^e\delta j^d \right) - \int_{\partial \Sigma} \left(
\delta{\mbox{\boldmath $Q$}} [\varphi]-\varphi \cdot {\mbox{\boldmath
$\Theta$}} \right)
\label{djem}
\end{eqnarray} 
Here $\Sigma$ is an arbitrary asymptotically flat hypersurface,
possibly possessing an inner boundary $\partial \Sigma$ (which may
be---but need not be---the horizon of a black hole), $\delta T_{ab}$
denotes the perturbed {\em non-electromagnetic} stress-energy tensor,
$\delta j^a$ denotes the perturbed electromagnetic charge-current
vector, and $A_a$ denotes the vector potential of the background in a
gauge compatible with the symmetries (i.e., ${\cal L}_t A_a = 0$ in
the stationary case, eq.(\ref{dmem}), and ${\cal L}_\varphi A_a = 0$ in
the axisymmetric case, eq.(\ref{djem})). The quantities
${\mbox{\boldmath $Q$}} $  and
${\mbox{\boldmath $\Theta$}}$ are given by eqs.(\ref{qgre}) and
(\ref{the123})-(\ref{defvd}) of section 2 below.

In section 3, we will give a proof of the physical process version of
the first law based on the above formulas. In section 4, we will
establish properties of the thermal atmosphere around a black hole
that follow from the assumptions stated above. The process of
quasi-statically lowering a box filled with matter towards a black
hole and then releasing it will be considered in section 5, and it
will be shown that the GSL holds in such processes. Our analysis of
the lowering process is compatible with (i.e., it does not conflict
with) the recent analysis of Shimomura and Mukohyama \cite{muko} for
charged, nonrotating black holes, but some of our arguments are quite
different from theirs, and we also clarify and generalize some aspects
of their derivation\footnote{In particular, in \cite{muko} the formula
for the gravitational force on the box is not derived, and it is
unclear at certain points whether their energy density, $\rho$,
includes (or should include) the electromagnetic interaction energy of
the charged matter with the background electromagnetic field. Also, a
proper justification for setting the chemical potential, $\mu$, to
zero on the horizon of the black hole was not given.}. We make some
concluding remarks in section 6.  In particular, we give an
independent argument that if the GSL could be violated in a
quasi-static lowering and release process involving a black hole, then
there should be a corresponding process involving a self-gravitating
system that does not contain a black hole in which the ordinary second
law would be violated\footnote{This thereby provides a response to
\cite{bek5} by showing that if the considerations of that paper could
lead to a violation of the GSL, then they also should give rise to a
violation of the ordinary second law.}. Finally, in the Appendix we
give a general derivation of the force needed to hold in place a box
containing charged matter in a stationary (but not necessarily static)
spacetime.

\section{First order variation of mass and angular momentum} \label{change}
In this section, we first consider the general issue of calculating
the first order variation of conserved quantities in a diffeomorphism
covariant theory of gravity in the case where the first order
perturbation is not required to satisfy the source free equations
(except near infinity). We will then specialize to the Einstein-Maxwell
case and derive formulas (\ref{dmem}) and (\ref{djem}) above.

Consider a diffeomorphism covariant theory in $n$-dimensions derived
from a Lagrangian $\mbox{\boldmath $L$}$, where  the dynamical fields consist of a
Lorentz signature metric $g_{ab}$ and other fields $\psi$. We will
follow the notational conventions of \cite{iyer1}, and, in particular,
we will collectively refer to $(g_{ab}, \psi)$ as $\phi$. The first
order variation of the Lagrangian can always be expressed in the form
\begin{eqnarray} 
\delta \mbox{\boldmath $L$} =\mbox{\boldmath $E$}  (\phi) \delta\phi+d\mbox{\boldmath $\Theta$}(\phi, \delta\phi)
\label{firval} 
\end{eqnarray}
where $\mbox{\boldmath $E$} (\phi)$ is locally constructed out of $\phi$ and its
derivatives and $\mbox{\boldmath $\Theta$}$ is locally constructed out of $\phi$,
$\delta\phi$ and their derivatives. The equations of motion then can
be read off as 
\begin{eqnarray} 
\mbox{\boldmath $E$} (\phi)=0
\label{eqofm} 
\end{eqnarray}
The symplectic current ($n-1$)-form $\mbox{\boldmath $\omega$}$ is defined by
\begin{eqnarray} 
\mbox{\boldmath $\omega$}(\phi,\delta_1 \phi, \delta_2 \phi)=\delta_1 \mbox{\boldmath $\Theta$}(\phi,
\delta_2 \phi) - \delta_2 \mbox{\boldmath $\Theta$}(\phi, \delta_1 \phi)
\label{omega} 
\end{eqnarray}

Let $\xi^a$ be any smooth vector field on the spacetime. We associate to
$\xi^a$ and $\phi$ a Noether current ($n-1$)-form, defined by
\begin{eqnarray}
\mbox{\boldmath ${\cal J}$}=\mbox{\boldmath $\Theta$} (\phi, {\cal L}_\xi \phi)-\xi\cdot\mbox{\boldmath $L$}  
\label{defjn}
\end{eqnarray}
where ``$\cdot$'' denotes contraction of the vector field $\xi^a$ 
into the first index of the differential form $\mbox{\boldmath $L$}$.
A simple calculation yields
\begin{eqnarray}
d\mbox{\boldmath ${\cal J}$}= -\mbox{\boldmath $E$}  (\phi)
{\cal L}_\xi\phi      
\label{dje}
\end{eqnarray}
It was proven in the Appendix of \cite{iyer2} there exists an 
($n-2$)-form $\mbox{\boldmath $Q$}$
(called the {\em Noether charge}), which is locally constructed from $\phi$,
$\xi^a$ and their derivatives, such that
\begin{eqnarray}
\mbox{\boldmath ${\cal J}$}[\xi] = d\mbox{\boldmath $Q$} [\xi] + \xi^a \mbox{\boldmath $C$}_a   
\label{jqca}
\end{eqnarray}
where $\mbox{\boldmath $C$}_a$ is an $(n-1)$-form (with an extra dual vector index)
which is locally constructed out of the dynamical fields and is such
that $\mbox{\boldmath $C$}_a=0$ when the equations of motion are satisfied.

Now suppose that the spacetime satisfies asymptotic conditions at
infinity corresponding to ``case I'' of \cite{zoupas} and that $\xi^a$
is an asymptotic symmetry. (``Case I'' of \cite{zoupas} is the case
where a true Hamiltonian corresponding to every asymptotic symmetry
exists, thereby giving rise to a conserved quantity, $H_\xi$,
associated with $\xi^a$. It includes the case of spacetimes that are
asymptotically flat at spatial infinity in general relativity.)  Let
$\delta \phi$ satisfy the linearized equations of motion, $\delta
\mbox{\boldmath $E$} (\phi) =0$, in a neighborhood of infinity, but not
necessarily throughout 
the spacetime. Then the variation of the conserved quantity, $\delta
H_\xi$ associated with $\xi^a$ is given by \cite{zoupas}
\begin{eqnarray} 
\delta H_\xi = \int_\infty (\delta\mbox{\boldmath $Q$}
[\xi]-\xi\cdot\mbox{\boldmath $\Theta$})
\label{ham} 
\end{eqnarray}
Here by ``$\int_\infty$'' we mean the following: Let $\Sigma$ be a
hypersurface in $M$ that extends smoothly to the boundary representing
infinity in the unphysical spacetime. We perform
the integral of eq.(\ref{ham}) over an $(n-2)$-surface in $\Sigma$ and
then take the limit as this $(n-2)$-surface goes to infinity along $\Sigma$; 
see \cite{zoupas} for further details.

Using Stokes' theorem, we may re-write eq.(\ref{ham}) as
\begin{eqnarray} 
\delta H_\xi = \int_\Sigma (\delta d\mbox{\boldmath $Q$}[\xi]- d (\xi\cdot\mbox{\boldmath $\Theta$}))+ \int_{\partial \Sigma} (\delta\mbox{\boldmath $Q$}
[\xi]-\xi\cdot\mbox{\boldmath $\Theta$})
\label{ham2} 
\end{eqnarray} 
where $\partial \Sigma$ denotes any ``interior boundary'' of $\Sigma$
(which would be empty if $\Sigma$ is a slice and there are no other
asymptotic regions, but we keep this term since we may wish to
terminate $\Sigma$ at, e.g., the event horizon of a black hole). Using
the identity \cite{iyer1} 
\begin{eqnarray} 
\delta\mbox{\boldmath ${\cal J}$} = \mbox{\boldmath $\omega$}(\phi,\delta \phi,{\cal L}_\xi\phi) + 
d(\xi\cdot\mbox{\boldmath $\Theta$})
\label{lomega} 
\end{eqnarray} 
we may eliminate the term $d(\xi\cdot\mbox{\boldmath $\Theta$})$ from eq.(\ref{ham2})
in favor of $\delta\mbox{\boldmath ${\cal J}$} $ and $\mbox{\boldmath
$\omega$}$.  

We now restrict
consideration to the case where $\xi^a$ is a Killing field of the
background spacetime and is also a symmetry of any background matter
fields $\psi$. Then $\mbox{\boldmath $\omega$}(\phi,\delta \phi,{\cal L}_\xi\phi) = 0$, so
eq.(\ref{ham2}) becomes
\begin{eqnarray}
\delta H_\xi &=&\int_\Sigma (\delta d\mbox{\boldmath $Q$} [\xi]-\delta
\mbox{\boldmath ${\cal J}$}[\xi]) 
+ \int_{\partial \Sigma} (\delta\mbox{\boldmath $Q$}[\xi]-\xi\cdot\mbox{\boldmath $\Theta$}) \nonumber \\
&=& -\int_\Sigma \xi^a\delta \mbox{\boldmath $C$}_a + \int_{\partial \Sigma} (\delta\mbox{\boldmath $Q$}
[\xi]-\xi\cdot\mbox{\boldmath $\Theta$})
\label{ham3}
\end{eqnarray}
where eq.(\ref{jqca}) was used in the last step. It is worth noting
that for an arbitrary perturbation, $\delta \phi$---i.e., $\delta
\phi$ need not satisfy the linearized field equations or have any
symmetries---of a solution, $\phi$, of the equations of motion
$\mbox{\boldmath $E$}  (\phi) = 0$ also satisfying ${\cal L}_\xi \phi = 0$, we have 
from eq.(\ref{jqca})
\begin{eqnarray}
d (\xi^a\delta \mbox{\boldmath $C$}_a) &=& d \delta \mbox{\boldmath ${\cal J}$}[\xi] - d^2 \delta \mbox{\boldmath $Q$}[\xi]  \nonumber \\
&=& 0
\label{cons}
\end{eqnarray}
where the variation of eq.(\ref{dje}) was used in the last step,
together with the fact that $\phi$ satisfies both $\mbox{\boldmath $E$} (\phi) = 0$
and ${\cal L}_\xi \phi = 0$. Thus, provided only that $\phi$
satisfies the equations of motion and ${\cal L}_\xi \phi = 0$, the current
\begin{eqnarray} 
\alpha^a = -\frac{1}{3!}\epsilon^{abcd} \delta C_{bcde} \xi^e 
\label{alpha} 
\end{eqnarray} 
is always conserved, $\nabla_a \alpha^a = 0$, where $\epsilon_{abcd}$ is the
metric compatible volume element of the background spacetime.

Eq.(\ref{ham3}) is our desired general formula for the first order
variation of conserved quantities. It holds for an arbitrary
diffeomorphism covariant theory of gravity derived from a Lagrangian
with an asymptotic region satisfying the conditions of ``case I'' of
\cite{zoupas}, provided only that $\xi^a$ is a symmetry of the
background spacetime (i.e., ${\cal L}_\xi \phi = 0$) and that $\delta
\phi$ satisfies the source-free linearized equations of motion near
infinity. Note that if $\delta \phi$ satisfies the linearized
equations of motion throughout the spacetime, then $\delta
\mbox{\boldmath $C$}_a =
0$ and the integral over $\Sigma$ in eq.(\ref{ham3}) vanishes. If, in
addition, $\Sigma$ has no interior boundary---i.e., if $\partial
\Sigma = \emptyset$---then eq.(\ref{ham3}) reduces to simply $\delta
H_\xi = 0$ (see \cite{sud}). On the other hand, if $\delta \phi$
satisfies the linearized equations of motion throughout the spacetime
but $\partial \Sigma$ is the bifurcation surface of the event horizon
of a stationary black hole, then eq.(\ref{ham3}) yields the general
form of the first law of black hole mechanics when $\xi^a$ is chosen
to be the horizon Killing field \cite{sud}, \cite{iyer1}. Our eq.(\ref{ham3})
generalizes these results by allowing $\delta \phi$ to fail to satisfy
the linearized equations except near infinity (i.e., by allowing
for the presence of sources for Einstein's equation as well as the
equations for the matter fields), as well as by allowing $\partial
\Sigma$ to be arbitrary.

We now specialize to Einstein-Maxwell theory, in order to obtain
explicit formulas for the variation of mass and angular momentum in
that case. The Einstein-Maxwell Lagrangian is
\begin{eqnarray}
\mbox{\boldmath $L$}=\frac{1}{16\pi}(\mbox{\boldmath $\epsilon$} R-\mbox{\boldmath$\epsilon$}
g^{ac}g^{bd}F_{ab}F_{cd} \label{lag})
\end{eqnarray}
where $\mbox{\boldmath $\epsilon$}$ is the volume element associated with the metric.
Computing the first order variation of $\mbox{\boldmath $L$} $, we obtain
\begin{eqnarray}
\delta\mbox{\boldmath $L$}=\frac{1}{16\pi}\mbox{\boldmath
$\epsilon$}(-G^{ab}+8\pi T_{\rm EM}^{ab}) \delta
g_{ab}+\frac{1}{4\pi}\mbox{\boldmath $\epsilon$}(\nabla_a
F^{ab})\delta A_b + d\mbox{\boldmath $\Theta$}
\label{dell}
\end{eqnarray}
where $T_{\rm EM}^{ab}$ is the stress energy tensor of the
electromagnetic field
\begin{eqnarray}
(T_{\rm EM})_{ab}=\frac{1}{4\pi}\left\{ F_{ac}F_b^{\
c}-\frac{1}{4}g_{ab}F_{de}F^{de}\right\}  
\label{deftem}
\end{eqnarray}
and
\begin{eqnarray}
\Theta_{abc}(\phi, \delta \phi)=\frac{1}{16\pi}\epsilon_{dabc}v^d      
\label{the123}
\end{eqnarray}
with
\begin{eqnarray} 
v_d=\nabla^b
\delta g_{db}-g^{ce}\nabla_d\delta g_{ce}-4F_d^{\ b}\delta A_b    \label{defvd}
\end{eqnarray} 
The (source free) Einstein-Maxwell equations can then be 
read off from equation (\ref{dell})
\begin{eqnarray}
G^{ab} - 8\pi T_{\rm EM}^{ab} &=& 0  \label{ein}  \\
\nabla_a F ^{ab}&=& 0  \label{max}
\end{eqnarray}

From (\ref{defjn}), we find that the
Noether current 3-form with respect to $\xi^a$ is given by
\begin{eqnarray}
{\cal J}_{abc}&=&dQ_{abc}^{\rm GR}+\frac{1}{16\pi}\epsilon_{dabc}(2G^d_{\ e}\xi^e
  +\xi^d F_{fg} F^{fg})    \nonumber  \\ 
 & -&\frac{1}{4\pi}\epsilon_{dabc} F^{df}(\xi^e\nabla_e A_f+A_e \nabla_f\xi^e)
\label{j3} 
\end{eqnarray}
where 
\begin{eqnarray*}
Q^{\rm GR}_{ab}=-\frac{1}{16\pi}\epsilon_{abcd}\nabla^c\xi^d
\end{eqnarray*}
Writing $\nabla_e A_b=F_{eb}+\nabla_b A_e$ in the last
term of (\ref{j3}) and differentiating by parts, we obtain
\begin{eqnarray}
{\cal J}_{abc} &=& dQ_{abc}^{\rm GR} + \frac{1}{8 \pi} \epsilon_{dabc}
(G^{de} -8\pi T_{\rm EM}^{de}) \xi_e -
\frac{1}{4\pi}\nabla_g(\epsilon_{dabc} F^{dg}A_e\xi^e) +
\frac{1}{4\pi}\epsilon_{dabc} A_e\xi^e\nabla_f F^{df} \nonumber \\ &=&
(dQ)_{abc} + \frac{1}{8 \pi}\epsilon_{dabc} (G^{de} -8\pi T_{\rm
EM}^{de}) \xi_e + \frac{1}{4\pi}\epsilon_{dabc} A_e\xi^e\nabla_f
F^{df}
\label{j5} 
\end{eqnarray}
where 
\begin{eqnarray}
Q_{ab}=-\frac{1}{16\pi}\epsilon_{abcd}\nabla^c\xi^d-\frac{1}{8\pi}\epsilon_{abcd}F^{cd}
A_e\xi^e   
\label{qgre}
\end{eqnarray} 
Eq.(\ref{j5}) is precisely of the required form (\ref{jqca}), so we may
identify $Q_{ab}$ as the Noether charge, and read off
$\mbox{\boldmath $C$}_a$ to be given by
\begin{eqnarray}
C_{bcda}=\frac{1}{8 \pi} \epsilon_{ebcd} ({G^e}_a - 8\pi {T^{{\rm EM}e}}_a) +\frac{1}{4\pi} \epsilon_{ebcd} A_a\nabla_f F^{ef}  
\label{ca}
\end{eqnarray}

Clearly, as is required, we have $\mbox{\boldmath $C$}_a = 0$ whenever the
source-free Einstein-Maxwell equations (\ref{ein}) and (\ref{max}) 
hold. When the
source-free Einstein-Maxwell equations do not hold, we write
\begin{eqnarray}
8\pi T^{de} = G^{de} - 8\pi T_{\rm EM}^{de}
\label{tne}
\end{eqnarray}
\begin{eqnarray}
4\pi j^{d} = \nabla_b F^{db}
\label{j}
\end{eqnarray}
Then $T_{ab}$ has the interpretation of being the {\em
non-electromagnetic} contribution to the stress energy tensor (i.e.,
$T^{de}=T_{\rm total}^{de}-T_{\rm EM}^{de}$) and $j^a$ is the
charge-current of the Maxwell sources. In terms of these sources, we have
\begin{eqnarray}
C_{bcda}=\epsilon_{ebcd}\left({T^e}_a + j^{e} A_a\right) 
\label{jdqj}
\end{eqnarray}

Now, let $(g_{ab}, A_a)$ be a solution of the source-free
Einstein-Maxwell equations (\ref{ein}) and (\ref{max}), and 
let $(\delta g_{ab},
\delta A_a)$ be a linearized perturbation which satisfies the
linearized Einstein-Maxwell equations with sources $\delta T_{ab}$ and
$\delta j^a$. Then, we have
\begin{eqnarray}
\delta C_{bcda}=\epsilon_{ebcd}\left(\delta
{T^e}_a + A_a \delta j^e  \right)  
\label{vajd}
\end{eqnarray}

Substituting (\ref{vajd}) into eq.(\ref{ham3}), we obtain the explicit 
formula
\begin{eqnarray}
\delta H_\xi
= -\int_\Sigma \epsilon_{ebcd}\left(\xi^a \delta
{T^e}_a + \xi^a A_a \delta j^e  \right) + \int_{\partial \Sigma} (\delta\mbox{\boldmath $Q$}
[\xi]-\xi\cdot\mbox{\boldmath $\Theta$})
\label{hamem}
\end{eqnarray}
where $Q_{ab}$ is given by eq.(\ref{qgre}) and $\Theta_{abc}$ is given
by eqs.(\ref{the123}) and (\ref{defvd}). Finally, choosing $\xi^a$ to
be an asymptotic time translation, $t^a$, and writing $M = H_t$, we
obtain eq.(\ref{dmem}) above, whereas choosing $\xi^a$ to
be an asymptotic rotation, $\varphi^a$, and writing $J = - H_\varphi$, we
obtain eq.(\ref{djem}) above.

\section{Physical process version of the first law of black hole mechanics} \label{physical}

Consider a classical, stationary black hole solution to the
Einstein-Maxwell equations (\ref{ein}) and (\ref{max}). Suppose we
perturb the black hole by dropping in some (possibly charged) matter.
If we assume that the black hole is not destroyed in this process
and that it eventually settles down to a stationary final state, we
can compute its change in mass and angular momentum using
eqs.(\ref{dmem}) and (\ref{djem}) above. We also can compute its
change in electric charge from the flux of charge-current through the
horizon, and we can compute its change in area using the Raychaudhuri
equation. In \cite{waldbook}, it was proven that eq.(\ref{pverf}) above
holds in the case where the unperturbed black hole has no
electromagnetic field, as is necessary for consistency with the first
law of black hole mechanics \cite{bch}, \cite{iyer1}. In this section,
we shall generalize this ``physical process'' version of the first law
to the case of charged black holes.

Let $(g_{ab}, A_a)$ be a solution to the source free
Einstein-Maxwell equations
(\ref{ein}) and (\ref{max}) corresponding to a stationary black hole.
Let
\begin{eqnarray}
\xi^a=t^a+\Omega_H \varphi^a  
\label{horkill}
\end{eqnarray} 
denote the horizon Killing field of this black hole
\cite{waldbook}. Let $\Sigma_0$ be an asymptotically flat hypersurface
which terminates on the event horizon ${\cal H}$ of the black hole. We
wish to consider initial data on $\Sigma_0$ for a linearized
perturbation $(\delta g_{ab}, \delta A_a)$ with matter sources $\delta
T_{ab}$ and $\delta j^a$ (see eqs.(\ref{tne}) and (\ref{j}) above). (We
emphasize that $\delta T_{ab}$ denotes the perturbation in the {\em
non-electromagnetic} contribution to the stress-energy tensor.) We
require that (i) $\delta T_{ab}$ and $\delta j^a$ vanish near infinity
and (ii) the initial data for $\delta g_{ab}$ and $\delta A_a$ (and,
hence, $\delta T_{ab}$ and $\delta j^a$) vanish in a neighborhood of
the horizon ${\cal H}$ on $\Sigma_0$---so that at the initial
``time'', $\Sigma_0$, the black hole is unperturbed. We assume that
all of the matter and charge eventually fall into the black hole, and
that the black hole eventually settles down to another stationary
black hole solution of the source free Einstein-Maxwell equations. Our
goal is to compute $\delta M$, $\delta J$, $\delta Q$, and $\delta A$
for the final state black hole and verify that eq.(\ref{pverf}) holds.

Since the perturbation vanishes near the internal boundary $\partial
\Sigma_0$ of the initial hypersurface, it follows immediately from
eq.(\ref{hamem}) (or equivalently, from eqs.(\ref{dmem}) and
(\ref{djem})) that the perturbed spacetime satisfies
\begin{eqnarray}
\delta M-\Omega_H\delta J=-\int_{\Sigma_0} \epsilon_{dabc}\left(\xi^e \delta
{T^d}_e + A_e \xi^e\delta j^d  \right) 
\label{demj}
\end{eqnarray}
Thus, the perturbed mass and angular momentum of the final black hole
will satisfy this relation. In terms of the current,
$\alpha$, 
\begin{eqnarray}
\alpha^a = \xi^b \delta {T^a}_b + A_b \xi^b \delta j^a
\label{alpha2}
\end{eqnarray}
(see eq.(\ref{alpha}) and eq.(\ref{vajd})), we have
\begin{eqnarray}
\delta M-\Omega_H\delta J=\int_{\Sigma_0} \alpha^d n_d \tilde{\epsilon}_{abc}
\label{demj2}
\end{eqnarray} 
where $n^a$ denotes the future-directed unit normal to $\Sigma_0$ and
$\tilde{\epsilon}_{abc}
= n^d \epsilon_{dabc}$. Using the
conservation of $\alpha^a$ and our assumption that all of the matter
eventually falls into the black hole, we can rewrite eq.(\ref{demj2})
as\footnote{In fact, eq.(\ref{demj3}) should hold for the black hole
final state even if all of the
matter and charge do not eventually fall into the black hole, with $\delta M$
and $\delta J$ being the perturbed mass and angular momentum of the
black hole---which no longer equal the perturbed mass and angular
momentum of the spacetime because of the presence of matter outside of
the black hole.}
\begin{eqnarray}
\delta M-\Omega_H\delta J=\int_{\cal H} \alpha^d k_d \tilde{\epsilon}_{abc}
\label{demj3}
\end{eqnarray} 
where $k^a$ is tangent to the affinely parametrized null geodesic generators 
of the event horizon, ${\cal H}$, of the unperturbed black hole and
$\tilde{\epsilon}_{abc}$ satisfies
\begin{eqnarray}
\frac{1}{4}\epsilon_{abcd}=- k_{[a}\tilde{\epsilon}_{bcd]}  
\label{epc}
\end{eqnarray}

The second term in $\alpha^a$ in eq.(\ref{alpha2}) yields a contribution
to the integral in (\ref{demj3}) of the form 
\begin{eqnarray}
I = - \int_{\cal H} \Phi_{\rm bh} \delta j^d k_d \tilde{\epsilon}_{abc} 
\label{I}
\end{eqnarray} 
where we have written
$\Phi_{\rm bh} = - (\xi^a A_a)|_{\cal H}$. However, $\Phi_{\rm bh}$ 
is constant over
the horizon of the black hole \cite{car}, as can be seen as follows.
We have
\begin{eqnarray}
\nabla_a(A_b \xi^b)&=& {\cal L}_\xi A_a+\xi^b(d A)_{ab} \nonumber \\
&=& {\cal L}_\xi A_a+\xi^b F_{ab} 
\label{nablaachi}
\end{eqnarray}
But ${\cal L}_\xi A_a$ vanishes since $\xi^b$ is symmetry of the
background solution. Furthermore, by the Raychauduri equation 
(see e.g. \cite{gr})
\begin{eqnarray}
\frac{d\theta}{dV}=-\frac{1}{2}\theta^2-\sigma_{ab}\sigma^{ab}-R_{ab}k^ak^b
\label{raych}  
\end{eqnarray}
(where $V$ denotes the affine parameter corresponding to $k^a$)
together with the fact that the expansion, $\theta$, and shear,
$\sigma_{ab}$, vanish in the stationary background, we have
\begin{eqnarray}
0=[T_{\rm EM}]_{ab} k^a k^b|_{\cal H}= F_{ac}{F_b}^c k^a k^b
\label{f1}
\end{eqnarray}
Consequently, $F_{ab} k^a$ is null, and since $F_{ab} k^a k^b = 0$ by
the antisymmetry of $F_{ab}$ it follows that $F_{ab} k^a$ is
proportional to $k_b$. Hence, the pullback of $F_{ab} k^a$ 
to ${\cal H}$ vanishes. Thus,
the pullback of $\nabla_a \Phi_{\rm bh}$ to ${\cal H}$ also vanishes, i.e., 
$\Phi_{\rm bh}$ is constant on ${\cal H}$, as we desired to show. 
Consequently, we have
\begin{eqnarray}
I &=& - \Phi_{\rm bh} \int_{\cal H} \delta j^d k_d \tilde{\epsilon}_{abc} \nonumber \\
&=& \Phi_{\rm bh} \delta Q 
\label{I2}
\end{eqnarray} 
where $\delta Q$ denotes the net flux of charge into the black hole.
From eqs. (\ref{demj3}), (\ref{alpha2}) and (\ref{I2}), we obtain
\begin{eqnarray}
\delta M-\Omega_H \delta J -\Phi_{\rm bh} \delta Q =
\int_{\cal H} \delta {T^d}_e \xi^e k_d   
\label{starone}
\end{eqnarray}

We now compute the change in area of the black hole. To simplify the
calculation, we use our diffeomorphism freedom in identifying the
perturbed spacetime with the background spacetime to make the null
geodesic generators of the event horizon of the perturbed black hole
coincide (as unparametrized curves) with the null geodesic generators
of the unperturbed stationary black hole. As a result of this gauge
choice, the perturbation in the location of the horizon vanishes, and
we have $\delta k^a \propto k^a$. The perturbed Raychaudhuri equation
(\ref{raych}) yields
\begin{eqnarray}
\frac{d(\delta \theta)}{dV}&=& -8\pi \delta ([T_{\rm total}]_{ab}k^a k^b) |_{\cal H}  \nonumber  \\
    &=& -8\pi \delta ([T_{\rm EM}]_{ab}) k^a k^b|_{\cal H} - 8\pi
    (\delta T_{ab}) k^ak^b|_{\cal H}  
\label{dtheta}
\end{eqnarray}
where we have used the fact that $[T_{\rm EM}]_{ab}k^a k^b|_{\cal H} = 0$
(see eq.(\ref{f1})) together with $\delta k^a \propto k^a$ to eliminate such
terms as $[T_{\rm EM}]_{ab} k^a \delta k^b$. However, we have
\begin{eqnarray}
\delta ([T_{\rm EM}]_{ab}) k^a k^b|_{\cal H} = (2 F_{ac} \delta {F_b}^c -\frac{1}{4}\delta g_{ab} F_{de} F^{de}-\frac{1}{2}
g_{ab} F_{de}\delta F^{de}) k^a k^b  
\label{rho3}
\end{eqnarray}
The last two terms vanish since $k^a$ is null in both the perturbed 
and unperturbed spacetimes. On the other hand, we showed above that 
$F_{ac} k^c \propto k_a$, so $F_{ac} \delta {F_b}^c k^a k^b \propto 
\delta F_{bc} k^b k^c = 0$ by antisymmetry of $\delta F_{bc}$. Thus, we 
obtain
\begin{eqnarray}
\frac{d(\delta \theta)}{dV} = -8\pi \delta T_{ab}k^ak^b|_{\cal H}  
\label{dtheta2}
\end{eqnarray}
A calculation identical to that given in \cite{waldbook} then yields
\begin{eqnarray}
\kappa \delta A=8\pi \int_{\cal H} \delta {T^d}_e \xi^e k_d   
\label{dA}
\end{eqnarray}
Substitution of this result into eq.(\ref{starone}) then yields 
eq.(\ref{pverf}), as we desired to show.

\section{The thermal atmosphere around a black hole} \label{thermal}

The remainder of this paper will be devoted to analyzing the validity
of the generalized second law (GSL) for processes in which some
(possibly charged) matter is quasi-statically lowered toward a
(possibly charged and rotating) black hole and then released. In this
section, we will state our assumptions about the thermal atmosphere
surrounding the black hole and derive certain properties of it.

An isolated black hole would continuously emit Hawking radiation to
infinity, and quantum fields around the black hole cannot come to
thermal equilibrium. However, thermal equilibrium should be possible
if the black hole is enclosed in a box. Nevertheless, even for an
uncharged and nonrotating black hole, this equilibrium will be
unstable unless the box enclosing the black hole is sufficiently small
\cite{gp}. For a rotating black hole, a more serious problem occurs:
There cannot exist a thermal equilibrium state of quantum fields
outside the black hole unless the black hole is enclosed in a
box that is sufficiently small that the horizon Killing field $\xi^a$,
eq.(\ref{horkill}), is timelike everywhere within the box
\cite{kw}. Similarly, for a charged black hole, thermal equilibrium of
a charged field outside the black hole would not be possible unless
the box is sufficiently small that the electrostatic potential
differences inside the box are insufficient to permit pair creation. In
the following, we will restrict consideration to the case where we
have a (possibly charged and rotating) black hole enclosed in a
sufficiently small box\footnote{Note that if the black hole were
enclosed in a box that was large enough for $\xi^a$ to become
spacelike and/or allow a large electrostatic potential difference, the
black hole would presumably simply lose angular momentum and/or discharge
until the quantum fields outside the black hole could reach a thermal
equilibrium state.} that the quantum fields outside of the black hole
are in a thermal equilibrium state with respect to the notion of
``time translations'' provided by the horizon Killing field $\xi^a$.

We shall refer to observers following orbits of $\xi^a$, i.e.,
observers with 4-velocity
\begin{eqnarray}
u^a = \xi^a/(-\xi^c \xi_c)^{1/2} 
\label{u}
\end{eqnarray}
as {\em stationary observers}.  We shall assume that the thermal
atmosphere admits a local thermodynamic description with respect to
stationary observers. There is a natural ground state of the quantum
field associated with $\xi^a$ \cite{kay}, \cite{waldbook}, which we
shall refer to as the {\em Boulware vacuum state}. It would be natural
for stationary observers to consider the non-electromagnetic
stress-energy tensor $T_{ab}$ and charge-current $j^a$ relative to the
Boulware vacuum, so we define
\begin{eqnarray}
\tilde{T}_{ab} = T_{ab} - (T_0)_{ab}
\label{Ttilde}
\end{eqnarray}
\begin{eqnarray}
\tilde{j}^a = j^a - j_0^a
\label{jtilde}
\end{eqnarray}
where $(T_0)_{ab}$ and $j_0^a$ denote, respectively, the (true,
renormalized) non-electromagnetic stress energy and charge-current of
the Boulware vacuum state.

It will be assumed that in thermal equilibrium, the
non-electromagnetic energy current $\tilde{T}_{ab} \xi^b$ and the
charge current $\tilde{j}^a$ relative to the Boulware vacuum state are
proportional to $\xi^a$. We also shall allow for the possibility that
other locally defined conserved currents $(\sigma_i)^a$ may
exist---such as, e.g., the number currents of various species of
particles that represent additional conservation laws beyond
conservation of charge. We also shall assume that in thermal
equilibrium the corresponding currents $(\tilde{\sigma}_i)^a$ measured
relative to the Boulware vacuum state are proportional to
$\xi^a$. Consequently, in thermal equilibrium, the above currents can
be characterized, respectively, by the energy density $\tilde{\rho} =
\tilde{T}_{ab} u^a u^b$, the charge density $\tilde{q} = - \tilde{j}_a
u^a$, and the quantities $\tilde{\lambda}_i = - (\tilde{\sigma}_i)_a
u^a$, all measured relative to the Boulware vacuum state.

It also will be assumed that in local thermal equilibrium the
stationary observers would assign an entropy current $\tilde{s}^a$ to
the thermal atmosphere, which also will be assumed to be proportional
to $\xi^a$, so that it also can be described by the entropy density
$\tilde{s} = - \tilde{s}_a u^a$, relative to the Boulware vacuum
state. We explicitly allow for the possibility that some
``renormalization'' of entropy may occur \cite{wa}, so that the ``true''
entropy density, $s$, of the thermal atmosphere is given by
\begin{eqnarray}
s = \tilde{s} + s_0
\label{stilde}
\end{eqnarray}
where $s_0$ is the (true) entropy density of the Boulware vacuum. It
would appear that such a renormaliztion of entropy must occur to avoid
a divergence in the contribution of the thermal atmosphere to the
total entropy due to $\tilde{s}$ becoming arbitrarily large near the horizon.

We will assume that $\tilde{\rho}$, $\tilde{q}$, and
$\tilde{\lambda}_i$ serve as ``state variables'' that characterize the local
thermodynamic state, so that the entropy density, $\tilde{s}$, can be
expressed as a function of these state variables
\begin{eqnarray}
\tilde{s} = \tilde{s}(\tilde{\rho}, \tilde{q}, \tilde{\lambda}_i)
\label{s}
\end{eqnarray}
However, it should be emphasized that we make no assumptions 
about the explicit 
functional form of $\tilde{s}$.

The renormalized non-electromagnetic stress energy, $T_{ab}$, and
charge-current, $j^a$, of the thermal atmosphere will, of course,
perturb the spacetime metric, $g_{ab}$, and electromagnetic field,
$A_a$, around the black hole. These perturbations of $g_{ab}$ and
$A_a$ will, in turn, affect the distribution and properties of the
thermal atmosphere. However, for a macroscopic black hole (i.e., a
black hole of mass much greater than the Planck mass), $T_{ab}$ and
$j^a$ will be small compared with scales set by the background
curvature, and we shall assume that they can be treated as linear
perturbations of the classical electrovac black hole spacetime. In
particular, we will assume that the formulas of section 2 apply for
the contribution of the thermal atmosphere to the total mass and
angular momentum of the spacetime. The effects of the perturbations of
$g_{ab}$ and $A_a$ on the thermal atmosphere would then be of second
and higher order, and therefore will be neglected.

In the following, we shall consider the effects of perturbing the
state of the thermal atmosphere to a nearby state that is locally in
thermal equilibrium. In accordance with the remarks in the previous
paragraph, we shall neglect the effects of the resulting perturbations
of $g_{ab}$ and $A_a$ when calculating the changes in the renormalized
$T_{ab}$ and $j^a$ of the thermal atmosphere caused by the
perturbation of its state. We shall similarly neglect the effects of
these perturbations of $g_{ab}$ and $A_a$ when calculating the changes
in $\tilde{T}_{ab}$ and $\tilde{j}^a$. In view of eqs.(\ref{Ttilde})
and (\ref{jtilde}), this additional assumption amounts to assuming that
the perturbations of $(T_0)_{ab}$ and $j_0^a$ are small compared with the
perturbations of $T_{ab}$ and $j^a$. Similar remarks apply to $(\sigma_i)^a$
and $(\tilde{\sigma}_i)^a$. Finally, we also shall neglect the effects of
the perturbations of $g_{ab}$ and $A_a$ when calculating the changes
in $s^a$ and $\tilde{s}^a$.

We now consider perturbing the state of the thermal atmosphere to a
nearby state that is locally in thermal equilibrium, characterized by
the state variables $\tilde{\rho} + \delta \tilde{\rho}$, $\tilde{q}
+ \delta \tilde{q}$, $\tilde{\lambda}_i + \delta \tilde{\lambda}_i$.
Variation of eq.(\ref{s}) yields the local form of the ordinary first
law of thermodynamics
\begin{eqnarray}
\delta \tilde{s} = \frac{1}{T} \delta \tilde{\rho} + \Psi \delta
\tilde{q} + \sum_i \gamma_i \delta \tilde{\lambda}_i
\label{ds}
\end{eqnarray}
where the temperature, $T$, and the potentials $\Psi$ and $\gamma_i$
are defined by appropriate partial derivatives of $\tilde{s}$ with respect to
the state variables. In a small volume, $V$, the locally measured
energy relative to the Boulware vacuum is $\tilde{U} = \tilde{\rho}
V$, and we similarly have $\tilde{Q} = \tilde{q} V$, and
$\tilde{\Lambda}_i = \tilde{\lambda}_i V$. Hence, we obtain
\begin{eqnarray}
\delta \tilde{S} &=& \delta (\tilde{s} V) \nonumber \\ 
&=& V[\frac{1}{T} \delta \tilde{\rho} + \Psi \delta \tilde{q} + \sum_i
\gamma_i \delta \tilde{\lambda}_i] + \tilde{s} \delta V \nonumber \\
&=& \frac{\delta \tilde{U}}{T} + \Psi \delta \tilde{Q} + \sum_i
\gamma_i \delta \tilde{\Lambda}_i + [\tilde{s} - \frac{\tilde{\rho}}{T}
- \Psi \tilde{q} - \sum_i \gamma_i \tilde{\lambda}_i] \delta V
\label{gd1}
\end{eqnarray}
We interpret the coefficient of $\delta V$ in this formula as
$\tilde{P}/T$, where $\tilde{P}$ denotes the pressure of the thermal
atmosphere relative to the Boulware vacuum.  We thereby obtain the
integrated form of the Gibbs-Duhem relationship for the thermal
atmosphere:
\begin{eqnarray}
\tilde{P} = T \tilde{s} - \tilde{\rho}
- T \Psi \tilde{q} - T \sum_i \gamma_i \tilde{\lambda}_i
\label{gd2}
\end{eqnarray}

Since only differences are taken in
eq.(\ref{ds}), the ground state contributions cancel out because, as
discussed in the previous paragraph, we neglect changes in the ground
state quantities. Therefore, eq.(\ref{ds}) also holds for the true,
renormalized quantities, i.e.,
\begin{eqnarray}
\delta {s} = \frac{1}{T} \delta {\rho} + \Psi \delta
{q} + \sum_i \gamma_i \delta {\lambda}_i
\label{ds2}
\end{eqnarray}
In the following, we will work with the true, renormalized quantities.

As discussed above, the contribution, 
\begin{eqnarray}
E \equiv M - \Omega_H J
\label{Edef}
\end{eqnarray}
of the thermal atmosphere to the total ``energy'' conjugate to $\xi^a$ of
the spacetime is given by (see eq.(\ref{hamem}))
\begin{eqnarray}
E = -\int_\Sigma \epsilon_{ebcd}\left(\xi^a {T^e}_a + \xi^a A_a j^e
\right)
\label{E}
\end{eqnarray}
Similarly, the contribution of the thermal atmosphere to the total
electric charge is
\begin{eqnarray}
Q = \int_\Sigma \epsilon_{ebcd} j^e 
\label{Q}
\end{eqnarray}
and its contribution to the globally conserved quantities,
$\Lambda_i$, associated with $(\sigma_i)^a$ is
\begin{eqnarray}
\Lambda_i = \int_\Sigma \epsilon_{ebcd} (\sigma_i)^e 
\label{Lam}
\end{eqnarray}
If the state of the thermal atmosphere is perturbed, variation of the
above formulas yields
\begin{eqnarray}
\delta E = \delta M - \Omega_H \delta J = -\int_\Sigma \epsilon_{ebcd}\left(\xi^a {\delta T^e}_a + \xi^a A_a \delta j^e
\right)
\label{dE}
\end{eqnarray}
\begin{eqnarray}
\delta Q = \int_\Sigma \epsilon_{ebcd} \delta j^e 
\label{dQ}
\end{eqnarray}
\begin{eqnarray}
\delta \Lambda_i = \int_\Sigma \epsilon_{ebcd} \delta (\sigma_i)^e 
\label{dLam}
\end{eqnarray}
However, since as discussed above, we have $\delta T_{ab} = \delta
\tilde{T}_{ab}$, $\delta j^a = \delta \tilde{j}^a$, and $\delta
(\sigma_i)^a = \delta (\tilde{\sigma}_i)^a$ and since the ``tilded''
currents have been assumed to be proportional to $\xi^a$, we may rewrite
eqs.(\ref{dE})-(\ref{dLam}) as
\begin{eqnarray}
\delta E = \delta M - \Omega_H \delta J = \int_\Xi (\chi \delta \rho - \xi^a A_a \delta q)
\label{dE2}
\end{eqnarray}
\begin{eqnarray}
\delta Q = \int_\Xi \delta q
\label{dQ2}
\end{eqnarray}
\begin{eqnarray}
\delta \Lambda_i = \int_\Xi \delta \lambda_i
\label{dLam2}
\end{eqnarray}
where $\Xi$ denotes the manifold of orbits \cite{ger} of $\xi^a$ with
the natural volume element $\bar{\epsilon}_{abc} = \epsilon_{dabc}
u^d$ on $\Xi$ understood. In eq.(\ref{dE2}), the ``redshift factor''
$\chi$ is defined by
\begin{eqnarray}
\chi \equiv (-\xi^a \xi_a)^{1/2}
\label{chi}
\end{eqnarray}

We now impose the assumption that the thermal atmosphere is in thermal
equilibrium with itself. More precisely, we assume that at fixed
total ``energy'', $E$, fixed total charge, $Q$, and fixed
$\Lambda_i$, the total entropy $S = \int_\Xi s$ is
maximum. In order for this to be the case, it is necessary for $S$ to be an 
extremum with respect to all first order variations that preserve the 
above constraints. We have
\begin{eqnarray}
\delta S &=& \int_\Xi \delta s \nonumber  \\
&=& \int_\Xi [\frac{1}{T} \delta \rho + \Psi \delta q + \sum_i \gamma_i 
\delta \lambda_i]
\label{varS} 
\end{eqnarray}
The necessary and sufficient conditions for $\delta S$ to vanish for
all variations $\delta \rho$, $\delta q$ and $\delta \lambda_i$
satisfying $\delta E = \delta Q = \delta \Lambda_i = 0$ can be
determined as follows. If we set $\delta q = \delta \lambda_i = 0$ (so
that we automatically satisfy $\delta Q = \delta \Lambda_i = 0$), we
see that by a suitable choice of $\delta \rho$ we can make $\delta S
\neq 0$ while preserving the remaining constraint $\delta E = 0$
unless the temperature, $T$, obeys the Tolman law
\begin{eqnarray}
T = T_0/\chi
\label{Tol}
\end{eqnarray}
where $T_0$ is a constant. The first term in eq.(\ref{varS}) can 
then be written as
\begin{eqnarray}
\int_\Xi \frac{1}{T} \delta \rho &=& \frac{1}{T_0} \int_\Xi \chi \delta \rho \nonumber \\
&=&  \frac{\delta E}{T_0} + \frac{1}{T_0} \int_\Xi \xi^a A_a \delta q
\label{rhoH} 
\end{eqnarray}
Substituting this into eq.(\ref{varS}), we find that by a suitable 
choice of $\delta q$ we can make $\delta S \neq 0$ unless $\Psi$ is 
of the form
\begin{eqnarray}
\Psi = - \frac{1}{T_0} (\xi^a A_a + \Phi_0) 
\label{Psi}
\end{eqnarray}
where $\Phi_0$ is a constant. Finally, it is easily seen that
extremization of $S$ under the constraints requires each $\gamma_i$ to
be constant\footnote{\label{chempot}If the conserved current $(\sigma_i)_a$
corresponds to a particle number current of a neutral particle, then
$\Lambda_i$ can be varied independently of $Q$ and the other particle
species. The chemical potential, $\mu_i$, of this species would then
be given in terms of $\gamma_i$ by $\mu_i = - T \gamma_i$. Taking
eq.(\ref{Tol}) into account, we see that the behavior of the chemical
potential corresponding to eq.(\ref{gam0}) is $\mu_i =
- T_0 \gamma_{0i}/\chi$.  If $(\sigma_i)_a$ is the particle number current
of a charged particle, then a variation of $\Lambda_i$ holding the
number of other particle species fixed requires a corresponding
variation of $Q$. The chemical potential for a species of charged
particles would be given by $\mu_i = (- T_0 \gamma_{0i} +e_i [\Phi_0 + \xi^a
A_a])/\chi$, where $e_i$ denotes the charge per particle of this
species.},
\begin{eqnarray}
\gamma_i = \gamma_{0i}
\label{gam0}
\end{eqnarray} 
It is easily seen that eqs.(\ref{Tol}), (\ref{Psi}), and
(\ref{gam0}) are also sufficient for $S$ to be an extremum. 

Finally, taking eqs.(\ref{Tol}), (\ref{Psi}), and (\ref{gam0}) into
account in eq.(\ref{varS}), we see that for a perturbation that does
not necessarily preserve $E$, $Q$, or $\Lambda_i$, the change in
the total entropy, $S$, of the thermal atmosphere is given by,
\begin{eqnarray}
\delta S &=& \int_\Xi [\frac{\chi}{T_0} \delta \rho - \frac{1}{T_0} (\xi^a A_a + \Phi_0)  \delta q + \sum_i \gamma_{0i} 
\delta \lambda_i] \nonumber  \\
&=& \frac{1}{T_0} (\delta E - \Phi_0 \delta Q) + \sum_i \gamma_{0i} 
\delta \Lambda_i] \nonumber  \\
&=& \frac{1}{T_0} (\delta M - \Omega_H \delta J - \Phi_0 \delta Q) 
+ \sum_i \gamma_{0i} \delta \Lambda_i]
\label{varS2} 
\end{eqnarray}
Eq.(\ref{varS2}) is the ``global form'' of the first law of
thermodynamics for the thermal atmosphere. Note that our above
analysis and results are applicable to any thermodynamic system that
is locally ``at rest'' with respect to a timelike Killing field
$\xi^a$, i.e., we did not use any special properties of the thermal
atmosphere to derive eqs.(\ref{Tol}), (\ref{Psi}), (\ref{gam0}), or
(\ref{varS2}), although in our formula for $E$, we did assume that the
non-electromagnetic stress-energy, $T_{ab}$, and charge-current,
$j^a$, could be treated as a linear perturbation of an electrovac
spacetime.

We now impose our assumption that the thermal atmosphere is in thermal
equilibrium with the black hole, i.e., that the total generalized
entropy, $S_{\rm C} \equiv S + S_{\rm bh}$, of the combined black hole/thermal
atmosphere system is at its maximum possible value for the given
values of the combined mass, $M_{\rm C} = M + M_{\rm bh}$, angular
momentum, $J_{\rm C} = J + J_{\rm bh}$, and charge, $Q_{\rm C} = Q +
Q_{\rm bh}$, of the total system. By the first law of black hole
mechanics, we have
\begin{eqnarray}
\delta S_{\rm bh} =  \frac{1}{T_H} (\delta M_{\rm bh} - \Omega_H \delta J_{\rm bh} - \Phi_{\rm bh} \delta Q_{\rm bh}) 
\label{varSbh} 
\end{eqnarray}
Comparing eqs.(\ref{varS2}) and (\ref{varSbh}), we see that $S_{\rm C}$
will be an extremum under interchange of mass, angular momentum, and
charge between the thermal atmosphere and the black hole if and only
if we have
\begin{eqnarray}
T_0 = T_H
\label{T0}
\end{eqnarray}
\begin{eqnarray}
\Phi_0 = \Phi_{\rm bh}
\label{Phi0}
\end{eqnarray}
and
\begin{eqnarray}
\gamma_{0i} = 0
\label{gam02}
\end{eqnarray}
Note that the vanishing of $\gamma_{0i}$ is essentially a consequence
of the ``no hair theorems'': Although for matter (including the
thermal atmosphere) there may be locally conserved currents
$(\sigma_i)^a$ aside from mass, angular momentum, and charge, there is
no global conservation law for these quantities when a black hole is
present, since the ``charges'' $\Lambda_i$ can fall into the black
hole, which retains no ``memory'' of them. If the charge $\Lambda_i$
corresponds to the number of particles of a particular species, then
the chemical potential for that species (see footnote \ref{chempot})
is given by
\begin{eqnarray}
\mu_i = e_i (\Phi_{\rm bh} - \Phi)/\chi
\label{cp}
\end{eqnarray}
It follows from eq.(\ref{cp}) that $\mu_i$ vanishes on the horizon of
the black hole, as had previously been claimed in \cite{muko}.

Taking into account the above relations, we find that the integrated
Gibbs-Duhem relation (\ref{gd2}) for the thermal atmosphere now takes
the form
\begin{eqnarray}
\chi \tilde{P} = T_H \tilde{s} - \chi \tilde{\rho}
- (\Phi - \Phi_{\rm bh}) \tilde{q}
\label{gd3}
\end{eqnarray}
We also note that the combined black hole/thermal atmosphere
system satisfies
\begin{eqnarray}
\delta S_{\rm C} &=& \frac{1}{T_H} (\delta M_{\rm C} - \Omega_H \delta
J_{\rm C} - \Phi_{\rm bh} \delta Q_{\rm C}) \nonumber \\
&=& \frac{1}{T_H} (\delta E_{\rm C} - \Phi_{\rm bh} \delta Q_{\rm C})
\label{varSC} 
\end{eqnarray}

\section{Lowering process}  \label{lowering}

In this section, we will consider a process in which a box containing
charged matter is quasi-statically lowered toward the black hole and
then dropped into the black hole or otherwise allowed to thermalize
with the black hole/thermal atmosphere system. The box will be assumed
to be ``perfectly insulating'', i.e., the walls will be assumed to be
perfectly reflecting with respect to the fields inside and outside of
the box. However, as will be discussed further below, the walls of the
box will necessarily radiate energy, charge, etc. into or out of the
box as the box is lowered \cite{unruhwald}. We make no assumptions
concerning the size, shape, or contents of the box other than that its
size is small compared with that of the black hole and that its
contents satisfy the ordinary thermodynamic laws (see below). In
particular, we do not make the ``thin box'' approximation
\cite{unruhwald}, nor do we even assume that the walls of the box are
rectangular in shape.

The process under consideration consists of two distinct stages: (1) A
quasi-static process in which the box is slowly lowered toward the
black hole. In this process, work may be done by an external agent, so
that the total energy contained in the box/black hole/thermal
atmosphere system may change. (2) A non-quasi-static process in which
the box is dropped into the black hole or otherwise destroyed, and the
system is allowed to thermalize. No change in the total energy or
charge of the complete system occurs in this stage. We will argue that
the total generalized entropy, $S'$, cannot decrease in either of these 
stages.

Consider, first, the quasi-static lowering process. At any given time
during this process, the total generalized entropy, $S'$ of the system
can be written as
\begin{eqnarray}
S' = S_{\rm B} + S_{\rm C} - S_{\rm D}
\label{totS}
\end{eqnarray}
Here, $S_{\rm B}$ denotes the total entropy contained in the box,
and $S_{\rm C}$ denotes the total entropy that the combined black
hole/thermal atmosphere system would have at the given values of $T_H$ and
$\Phi_{\rm bh}$ if the box were not present. Finally, $S_{\rm D}$ denotes
the entropy of the displaced thermal atmosphere, i.e., the entropy
that would have been contained in the thermal atmosphere (at the given
values of $T_H$ and $\Phi_{\rm bh}$) within the region occupied by the
box.

We now focus attention on $S_{\rm B}$. It is instructive to consider,
first, the case where the box initially is ``empty'', i.e., the
initial state of the fields inside the box is the natural
vacuum/ground state relative to the notion of time translations
defined by $\xi^a$ (see \cite{kay}). Then, we claim that if the
lowering process is sufficiently slow, no ``particle creation'' will
occur as the box is lowered, and the box will remain in its ground
state. In other words, an empty box will remain empty as it is
quasi-statically lowered. By definition, the Casimir energy of the box
is the difference between the energy contained in the empty box and
the energy that would be contained in the Boulware vacuum in the
region of space occupied by the box. (Recall here that by ``energy'',
$E$, we mean the conserved quantity conjugate to the horizon Killing
field $\xi^a$ rather than the asymptotic time translation $t^a$, i.e.,
$E = M - \Omega_H J$ (see eq.(\ref{Edef}) above).) If we neglect
possible changes in the Casimir energy of the box as it is lowered, it
follows that in the lowering process for an empty box, we have
\begin{eqnarray}
\Delta E_{\rm B} = \Delta E_0
\label{DE}
\end{eqnarray}
where $\Delta E_{\rm B}$ denotes the difference in the energies
contained in the box at two stages of the lowering process, and
$\Delta E_0$ denotes the difference in the Boulware vacuum energies in
the corresponding volumes. Here we have written ``$\Delta$'' rather
than ``$\delta$'' to emphasize that we are taking differences of
quantities associated with different regions of space rather than
differences of quantities associated with the same region of space, as
in the previous section.  

Similarly, the total charge $Q_{\rm B}$ within the empty box satisfies
\begin{eqnarray}
\Delta Q_{\rm B} = \Delta Q_0
\label{DQ}
\end{eqnarray}
It should be noted that changes in the charge contained within the box
can occur only as a result of radiation by the walls of the
box. However, this radiation by the walls of the box is independent of
the contents of the box. Therefore, in all cases (i.e., whether or not
the box is empty), the change in the charge of the box as it is
lowered is given by eq.(\ref{DQ}). However, eq.(\ref{DE}) holds only
for the empty box, since the total energy contained within the box can
vary due to redshifting of the energy of the contents of the box as
well as by radiation by the walls of the box.

Now suppose that the box is initially filled with (possibly charged)
matter that is in thermal equilibrium. As the box is lowered, the
matter will, in general, redistribute itself within the box due to the
changing electromagnetic and gravitational fields. However, if the box
is lowered sufficiently slowly, the matter will remain in thermal
equilibrium as it is lowered. Observers inside of the box will view
the process as being isentropic for the same reason as slow variations
of parameters in the Hamiltonian result in isentropic processes in
flat spacetime physics. Thus, the entropy above the ground state
must remain constant\footnote{This result also can be derived from
energy balance considerations, as outlined in footnote 2 of
\cite{mark}. However, this argument assumes that no ``extra energy''
(i.e., ``heat'') is fed into or taken out of the box as it is lowered,
and thus, in essence, {\it assumes} that the lowering process is
isentropic.} as the box is lowered. Taking into account the
possibility that the ground state entropy is nonzero due to
``renormalization'' as discussed in the previous section, we see that
in a slow lowering process where the matter is initially in thermal
equilibrium, we have $\Delta S_{\rm B} = \Delta S_0$, where $S_0$
denotes the entropy of the Boulware vacuum in the region occupied by
the box. Equivalently, we have $\Delta \tilde{S}_{\rm B} = 0$, where in
accord with the notation of the previous section, $\tilde{S}_{\rm B} \equiv
S_{\rm B} - S_0$.

Finally, suppose now that the box is initially filled with arbitrary
matter, not necessarily in thermal equilibrium. Then, as the box is
lowered, the matter may (partially or fully) thermalize. Observers
inside the box will see an increase---or, at least, a
non-decrease---of entropy relative to the ground state during the
lowering process for the same reason that the ordinary second law of
thermodynamics holds in flat spacetime physics. Consequently, we
conclude that whatever is initially placed inside the box, we have
\begin{eqnarray}
\Delta \tilde{S}_{\rm B} \equiv \Delta S_{\rm B} - \Delta S_0 \geq 0
\label{DSB}
\end{eqnarray}
during the lowering process.

We now turn our attention to the calculation of the change, $\Delta
S_{\rm C}$, in $S_{\rm C}$ during the lowering process. By
eq.(\ref{varSC}), we have
\begin{eqnarray}
\Delta S_{\rm C}=\delta S_{\rm C} = \frac{1}{T_H} (\delta E_{\rm C} -
\Phi_{\rm bh} \delta Q_{\rm C})
\label{varSC2} 
\end{eqnarray}
so we need to calculate the energy, $\delta E_{\rm C}$, and charge,
$\delta Q_{\rm C}$ delivered to the black hole/thermal atmosphere
system during the lowering process. The total charge of the entire
system is
\begin{eqnarray}
Q' = Q_{\rm B} +  Q_{\rm C} - Q_{\rm D}
\label{Q'} 
\end{eqnarray}
so by conservation of charge, we have
\begin{eqnarray}
\delta Q_{\rm C} = - \Delta Q_{\rm B} + \Delta Q_{\rm D}
\label{Qcons} 
\end{eqnarray}
We already found above that $\Delta Q_{\rm B} =
\Delta Q_0$ (see eq.(\ref{DQ})), so we have
\begin{eqnarray}
\delta Q_{\rm C} = - \Delta Q_0 + \Delta Q_{\rm D} = \Delta \tilde{Q}_{\rm D}
\label{dQC} 
\end{eqnarray}

To calculate $\delta E_{\rm C}$, we note that this quantity cannot
depend upon what is placed inside the box during the lowering process,
so it suffices to restrict attention to the case where the box is
empty. In that case, the change in total energy inside the box is
given by eq.(\ref{DE}), so by conservation of energy, we have
\begin{eqnarray}
\delta E_{\rm C} = W - \Delta E_0 + \Delta E_{\rm D} = W + \Delta
\tilde{E}_{\rm D}
\label{dEC} 
\end{eqnarray}
where $W$ denotes the work done by the external agent during the
lowering process. We calculate $W$ as follows. As shown in the
Appendix, the net force $F^a$ exerted
by an external agent at redshift $\chi_0$
on a stationary box whose size is much smaller than the scales set by
curvature is given by\footnote{Here we neglect any contributions
of the Casimir energy and/or the walls of the box to
$\tilde{\rho}$. These will not, in any case, affect the energy
delivered to the black hole/thermal atmosphere system, provided that
their locally measured stress-energy remains constant during the
lowering process, so that eq.(\ref{DE}) holds.}
\begin{eqnarray}
\chi_0 F^a = e^a \int_{\partial {\rm B}} \chi \tilde{P} e_b n^b dA
\label{F}  
\end{eqnarray}
Here $e^a$ denotes the unit ``upward pointing'' tangent to the string
(defined over the volume of the box via parallel transport---see the
Appendix for details) and $n^a$ is the unit outward pointing normal to
the surface, $\partial {\rm B}$, of the box in the manifold of orbits
of $\xi^a$.  If the box is lowered, the work done by the external
agent during the lowering process is
\begin{eqnarray}
W = - \chi_0 \int (F^a e_a)|_l dl = - \int \chi \tilde{P} e_a n^a dAdl
\label{W}  
\end{eqnarray}
where $l$ denotes proper length along the path of lowering in the
manifold of orbits. Equation (\ref{W}) is easiest to analyze in the
case of a rectangular box with ``top'' and ``bottom'' faces
perpendicular to $e^a$. In that case, eq.(\ref{W}) corresponds to
performing a volume integral of $\chi \tilde{P}$ over the spatial
region swept out by the top face of the box, and subtracting from it
the similar integral over the spatial region swept out by the bottom
face. The result is
\begin{eqnarray}
W = \Delta \int_{\rm B} \chi \tilde{P} dV
\label{W2}  
\end{eqnarray}
where the integral is taken over the volume of the box, and $\Delta$
denotes the difference between the final and initial values of this
integral in the lowering process. By a similar argument, it is not
difficult to see that eq.(\ref{W2}) remains valid for a box of
arbitrary shape.

We now apply the integrated Gibbs-Duhem relation (\ref{gd3}) for the 
thermal atmosphere. We thereby obtain
\begin{eqnarray}
W &=& \Delta \int [T_H \tilde{s} - \chi \tilde{\rho} - (\Phi -
\Phi_{\rm bh}) \tilde{q}] \nonumber \\ &=& T_H \Delta \tilde{S}_{\rm D} -
\Delta \tilde{E}_{\rm D} + \Phi_{\rm bh} \Delta \tilde{Q}_{\rm D}
\label{W3}
\end{eqnarray}
where the subscript $\rm D$ denotes quantities associated with the
displaced thermal atmosphere and eq.(\ref{dE2}) was used. Combining
eqs.(\ref{varSC2}), (\ref{dQC}), (\ref{dEC}), and (\ref{W3}), we obtain
\begin{eqnarray}
\Delta S_{\rm C} &=& \frac{1}{T_H}[\delta E_{\rm C} - \Phi_{\rm bh}
\delta Q_{\rm C}] \nonumber \\
&=& \frac{1}{T_H}[W + \Delta \tilde{E}_{\rm D} - \Phi_{\rm bh}
\delta Q_{\rm C}] \nonumber \\
&=& \Delta \tilde{S}_{\rm D}
\label{varSC3} 
\end{eqnarray}
Combining this equation with eqs.(\ref{totS}) and (\ref{DSB}), we obtain
\begin{eqnarray}
\Delta S' = \Delta \tilde{S}_{\rm B} \geq 0
\label{glslow} 
\end{eqnarray}
which shows that the generalized second law holds during the lowering
process.

Now, consider the ``dropping process'', i.e., we suppose that---after
completion of the above lowering process---the box is released and
allowed to fall into the black hole (or that the box is destroyed and
its contents are allowed thermalize with the black hole/thermal
atmosphere system). We assume that at the end of this process, the
final state of the system is that of a black hole in equilibrium with
its thermal atmosphere. Although the ``dropping process'' is a highly
nonequilibrium process and we cannot analyze the time evolution of the
total entropy during this process, it is clear that if the second assumption
of section 1 holds, the total generalized entropy, $S'$, cannot
decrease in this process. Namely, during the ``dropping process'' the
total mass, angular momentum, and charge of the system remain
constant. However, assumption (2) asserts that the final state
maximizes the total generalized entropy at the given values of total
mass, angular momentum, and charge. Therefore, the initial state could
not have had more generalized entropy than the final state.

Consequently, under the assumptions stated in section 1 as well as our
assumptions about the thermodynamic properties of the thermal
atmosphere stated in section 4, the GSL cannot be violated in any
quasi-static lowering/dropping process.

\section{Concluding remarks on the validity of the GSL}

In this paper, we have established the validity of the GSL in
arbitrary quasi-static lowering/dropping processes for charged and
rotating black holes, without the need to assume any entropy bounds
for matter. Our analysis depends, primarily, only on the two very
general assumptions stated in section 1. It thereby generalizes and
simplifies previous analyses of such processes.

However, it should be noted that during the course of our analysis, we
made several simplifying assumptions concerning the thermodynamic
properties of the thermal atmosphere. In particular, it was assumed
that stationary observers would (i) assign a locally homogeneous
entropy density $\tilde{s}$ to the thermal atmosphere that is valid on
all relevant scales and (ii) that $\tilde{s}$ is a function only of
$\tilde{\rho}$, $\tilde{q}$, and $\tilde{\lambda}_i$. The first
assumption need not be valid if one considers boxes whose size is
small compared with the wavelength of the ambient thermal atmosphere
(see \cite{bek5}). The second assumption could fail because the formula for
$\tilde{s}$ might also depend nontrivially upon ``location in the
gravitational field'' (e.g., depend upon the local value of the
curvature and/or derivatives of $\xi^a$).

In this concluding section, we wish to argue that if the breakdown of
any such simplifying assumptions about the thermal atmosphere were to
allow one to violate the GSL, they also would allow one to violate the
ordinary second law. Specifically, suppose that---on account of, say,
the breakdown of assumptions (i) or (ii) of the previous
paragraph---it were possible to violate the GSL in a quasi-static
lowering/dropping process. Since the validity of the GSL during the
``dropping'' phase does not depend upon any simplifying assumptions
about the thermal atmosphere, it follows that a violation of the GSL
must occur in the lowering phase. Now, in the lowering phase, the box
will stay some finite distance, $\epsilon$, outside of the horizon of
the black hole. Then, it should be possible, in principle, to
construct a (charged and rotating) shell with a perfectly reflecting
surface whose exterior gravitational and electromagnetic fields are
coincide with those of the black hole at distances greater than
$\epsilon$ from the horizon. We can enclose this shell in a cavity of
the same size as used for the black hole and then fill this cavity
with ``real'' thermal radiation in such a way that its temperature is
$T_H/\chi$, its potential $\Phi_0$ is equal to $\Phi_{\rm bh}$ and the
potentials $\gamma_{0i}$ are equal to zero. (This can be accomplished
by supplying the atmosphere with the appropriate amounts of energy,
charge, and other conserved quantities.) There may be slight
differences between the properties of the thermal atmosphere of the
black hole and those of the ``real'' atmosphere around the shell that
we have put in ``by hand'' on account of slight differences between
the ground states and modes in the two cases. However, these
differences can be made arbitrarily small by choosing the radius of
the shell to be arbitrarily close to the horizon radius of the black
hole.

Now, the analysis of lowering process given in the previous section
applies without change to a lowering process where the black
hole/thermal atmosphere system is replaced by the shell surrounded by
a ``real'' atmosphere, provided that the subscript ``$\rm C$'' is now
interpreted as referring to the ``real'' atmosphere around the
shell. The values of $S_{\rm C}$, $E_{\rm C}$, etc. may be very
different for the ``real'' atmosphere around the shell as compared
with the black hole/thermal atmosphere system, but the variations of
these quantities during the lowering process will be the same
(provided that the radius of the shell is sufficiently close to the
horizon radius of the black hole). It follows that if a lowering
process that decreases the total generalized entropy can be done in
the black hole case, a corresponding lowering process in the case of
the shell will decrease the total ordinary entropy. Thus, if the GSL
can be violated, then a corresponding process will violate the
ordinary second law.

\bigskip
\bigskip

\noindent
{\bf Acknowledgements}

This research was supported in part by NSF grants PHY95-14726 and
PHY00-90138 to the University of Chicago.

\appendix\section{Appendix: Force exerted on a stationary box}

In this Appendix, we consider a stationary (but not necessarily
static) spacetime, with timelike Killing vector field $\xi^a$. We
consider a stationary box in this spacetime which is held in place by
an agent who holds a massless string that is connected to the box (see
Fig. 1). The box may be of arbitrary shape and may contain charged
matter. An external electromagnetic field may be present and there
also may be additional matter outside of the box (which may exert a
``buoyancy force'' on the box). We wish to calculate the force that
the agent must exert on the ``far end'' of the string in order to hold
the box in place under the following assumptions:

We assume that the world sheet of the string is invariant under
$\xi^a$, and has stress-energy of the form
\begin{eqnarray}
T_{\rm S}^{ab}=P e^a e^b  
\label{strs}
\end{eqnarray}
where $e^a$ is a unit vector that is tangent to
the world sheet of the string and is orthogonal to $\xi^a$. (We choose
the direction of $e^a$ to point ``towards the agent'', i.e., ``away
from the box''.) This stress tensor corresponds to a massless string, which we
consider for simplicity; it is straightforward to allow the string to
have mass, but then the weight of the string would contribute to the
force exerted by the external agent. The ``$P$'' in eq.(\ref{strs}) is
understood to be proportional to a delta-function on the world sheet
of the string. We also assume that the string does not contain any
electromagnetic charge or current.

\begin{figure}[t]

\resizebox{\textwidth}{!}
{\includegraphics[-5.5in,0in][7.5in,8in]{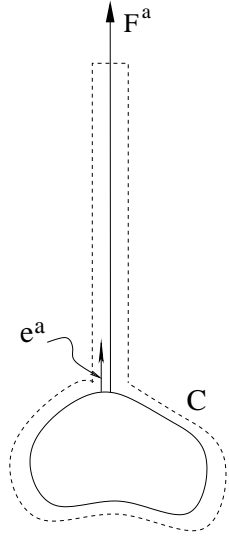}}

\caption{In a stationary spacetime, a box is held in place by an agent
who holds a massless string connected to the box. The surface, $C$,
enclosing the box and string is represented by the dotted line.}
\end{figure}

We decompose the stress-energy tensor of everything---including the
contents of the box, the walls of the box, the string, and the matter
outside of the box---into its electromagnetic and non-electromagnetic
parts
\begin{eqnarray}
T_{\rm total}^{ab}= T^{ab}+T_{\rm EM}^{ab}  
\label{strete}
\end{eqnarray}
where $T_{\rm EM}^{ab}$ denotes the stress tensor of the electromagnetic
field (see eq.(\ref{deftem})).  We assume that the electromagnetic
field is stationary
\begin{eqnarray}
{\cal L}_\xi A_a=0 
\label{lA}
\end{eqnarray}
This implies, of course, that $T_{\rm EM}^{ab}$ and the electromagnetic
charge-current vector, $j^a$, also are Lie derived by $\xi^a$. We
further assume that $j^a$ takes the form
\begin{eqnarray}
j^a&=&q u^a=\frac{q}{\chi}\xi^a
\label{ju}
\end{eqnarray}
where $u^a = \xi^a/\chi$ with $\chi = (-\xi^a \xi_a)^\frac{1}{2}$,
i.e., we assume that the charges are ``at rest'' (no current flow)
with respect to the stationary observers.  Similarly, we assume that
the total non-electromagnetic stress-energy tensor, $T_{ab}$, is
stationary
\begin{eqnarray}
{\cal L}_\xi T_{ab}=0 
\label{ltab}
\end{eqnarray}
and we further assume that $T_{ab}$ takes the form
\begin{eqnarray}
T^{ab}&=&\rho u^a u^b+ t^{ab}   
\label{noest}
\end{eqnarray}
with $t_{ab} u^a = 0$, i.e., we assume that the non-electromagnetic
stress-energy tensor has no ``time-space'' components (i.e., no
momentum density) relative to the stationary observers. Note that our
assumptions concerning $T^{ab}$ are compatible with our assumed form
of the stress-energy tensor of the string, eq.(\ref{strs}), which is
included in $T^{ab}$.

It is convenient to work on the manifold, $\Xi$, of orbits of $\xi^a$
(see \cite{ger}). All tensor fields on $M$ that are Lie derived by
$\xi^a$ and have all indices perpendicular to $\xi^a$ have a natural
projection to $\Xi$, and we will not distinguish in our notation such
spacetime tensors from their projections to $\Xi$. In particular,
$\Xi$ naturally acquires a Riemannian metric $h_{ab}$ given by
\begin{eqnarray}
h_{ab}=g_{ab}+ u_a u_b  
\label{deha}
\end{eqnarray}
We denote by $D_a$ the derivative operator on $\Xi$ associated with
$h_{ab}$. Our final assumption is that the size of the box is small
compared with the scales of curvature in the manifold of orbits. 

The string stress-energy $T_{\rm S}^{ab}$, eq.(\ref{strs}), must be
conserved everywhere except at the endpoints of the string. This
implies that $e^a$ must be a geodesic in spacetime, $e^b \nabla_b e^a
= 0$. It follows immediately that $e^b D_b e^a = 0$, i.e., the
projection of the string to the manifold of orbits, $\Xi$, is a
geodesic in the manifold of orbits. We now choose a surface $C$ in
$\Xi$ which encloses the box and string in the manner shown in
Fig. 1. We extend the definition of $e^a$ to the interior of $C$ by
parallel transport (with respect to $D_a$) along geodesics (with
respect to $D_a$) starting from the point at which the string is
attached to the box. (Note that since the size of the box has been
assumed to be small compared with scales set by curvature, parallel
transport over the box will be essentially path independent in any
case.)

Conservation of the total stress energy, eq.(\ref{strete}), yields
\begin{eqnarray}
0&=&\nabla_b T_{\rm total}^{ab} \nonumber \\ 
&=& \nabla _b (\rho u^au^b) +\nabla_b t^{ab}+\nabla_b T_{\rm EM}^{ab} \nonumber \\
&=& \rho u^b \nabla _b u^a +\nabla_b t^{ab} - F^{ab} j_b
\label{consl}
\end{eqnarray}
where we have used $u^b \nabla_b \rho = 0$ and $\nabla_b u^b = 0$ in
the last line. Since $u^a = \xi^a/\chi$, we obtain
\begin{eqnarray}
u^b \nabla_b u^a = \frac{1}{\chi} D^a \chi  
\label{acc}
\end{eqnarray}
On the other hand,
\begin{eqnarray}
F_{ab} j^b&=& \left(\nabla_a A_b-\nabla_b A_a\right) \frac{q}{\chi}\xi^b
\nonumber \\
&=& \frac{q}{\chi}[-{\cal L}_\xi A_a +\nabla_a(A_b\xi^b)] \nonumber \\
&=& - \frac{q}{\chi} D_a \Phi
\label{rimax}
\end{eqnarray}
where $\Phi \equiv - A_a \xi^a$.
Thus, we obtain
\begin{eqnarray}
0 = \frac{\rho}{\chi} D^a \chi +\nabla_b t^{ab} + \frac{q}{\chi} D^a \Phi
\label{consl2}
\end{eqnarray}

We now contract this equation with $e_a$, using
\begin{eqnarray}
e_a \nabla_b t^{ab} &=& \nabla_b (e_a t^{ab}) - t^{ab} \nabla_b e_a
\nonumber \\
&=& \frac{1}{\chi} D_b (\chi e_a t^{ab}) - t^{ab} D_b e_a
\label{divt}
\end{eqnarray}
Here we have used the identity
\begin{eqnarray}
\nabla_bv^b=\frac{1}{\chi}D_b(\chi v^b)  
\label{idenn}
\end{eqnarray}
that holds for any vector field $v^a$ in the class that projects to
$\Xi$, and we also changed $\nabla_b$ to $D_b$ in the second term
since $t^{ab}$ has both indices perpendicular to $\xi^a$. We thus obtain,
\begin{eqnarray}
0=\rho e^a D_a\chi+q e^a D_a \Phi + D_b[\chi t^{ab} e_a]-\chi t^{ab} D_b e_a 
\label{leim}
\end{eqnarray}

By construction, $D_b e_a$ vanishes at the point where the string is
attached to the box. If the geometry of $\Xi$ were flat, then $D_b
e_a$ would vanish identically throughout the box. Since the geometry
of $\Xi$ is not flat, $D_b e_a$ is, in general, nonvanishing. However, its
magnitude is bounded by the size of the box times the curvature
of $\Xi$. Therefore, for a box whose size is small compared with the
scales set by the curvature of $\Xi$, the last term in eq.(\ref{leim})
will be negligible compared with the other terms in that
equation. Therefore, we shall drop this term.

Integrating the remaining three terms in eq.(\ref{leim}) over
the volume, $V$, enclosed by $C$ and using Gauss's law, we obtain
\begin{eqnarray}
0 = \int_V \left[ \rho e^a D_a \chi+q e^a D_a\Phi\right] +
\int_{C}\chi t^{ab} e_a n_b dS  
\label{f111}
\end{eqnarray}
where the natural volume elements (with respect to $h_{ab}$) on $V$
and $C$ are understood, and $n^a$ is the unit, outward pointing normal
to $C$. We now ``shrink $C$ down'' so that it just barely encloses the
box and string. In this limit, the volume integral receives no
contribution from the string (since we assume that $\rho = q = 0$ on
the string), and the surface term also receives no contribution from
the portion surrounding the string (since the area of this portion
goes to zero), except for the contribution $\int \chi P = \chi_0 \int
P$ arising from the endpoint of the string held by the external agent,
where $\chi_0$ denotes the value of $\chi$ at this endpoint. Since
$F^a = - e^a \int P$ is just the force that the external agent must
exert to counterbalance the tension/pressure of the string and thereby
hold the box in place, we obtain the desired general expression for
the force needed to hold the box in place,
\begin{eqnarray}
\chi_0 F^a = e^a \left(\int_B \left[\rho e^b D_b \chi + q e^b D_b\Phi\right] +
\int_{\partial B}\chi t^{cb} e_c n_b dS \right) 
\label{f11}
\end{eqnarray}
where the volume integral is now taken over the box, $B$, and the
surface integral is taken over the boundary of the box, $\partial B$.
The first term in the volume integral can be interpreted as the
``weight'' of the contents of the box. Note that $\rho$ includes only
the non-electromagnetic energy density, i.e., neither the
electromagnetic self-energy of charges within the box nor their
interaction energy with external electromagnetic fields contribute to
the first term. The second term is just the Lorentz force on the
charge distribution in the box (including ``self-force'' effects). The
final term corresponds to the buoyancy force exerted on the box by the
matter surrounding the box.

We now specialize this result to the case of a box held near a black
hole, which is surrounded by the thermal atmosphere of the black
hole. In this case, we take $\xi^a$ to be the horizon Killing field,
eq.(\ref{horkill}). However, there is no reason to expect the true,
renormalized charge-current, $j^a$, will be of the form (\ref{ju}) nor
do we expect the renormalized nonelectromagnetic stress-energy tensor,
$T^{ab}$, to be of the form (\ref{noest}), since the charge-current
and stress-energy of the Boulware vacuum would not be expected to have
this form. However, it seems reasonable to expect that the
differences, $\tilde{j}^a$ and $\tilde{T}^{ab}$, between the true
charge-current and stress-energy and those of the Boulware vacuum (see
eqs.(\ref{Ttilde}) and (\ref{jtilde}) above) will have this form. Now,
the total stress-energy, $(T_0)^{ab}_{\rm total}$, of the Boulware
vacuum must be conserved, so $\tilde{T}^{ab}_{\rm total}$ also is
conserved. If treat both $T^{ab}_{\rm EM}$ and $(T_0)^{ab}_{\rm EM}$
as small perturbations of the electromagnetic stress-energy tensor of
the black hole (so that only linear terms in the deviation from the
background black hole electromagnetic stress-energy are kept), then a
repetition of the steps in the above derivation shows that in this
case, eq.(\ref{f11}) continues to hold, provided only that $\rho$,
$q$, and $t^{ab}$ are replaced by their ``tilded'' counterparts, i.e.,
we have
\begin{eqnarray}
\chi_0 F =  \int_B \left[\tilde{\rho} e^a D_a \chi + \tilde{q} e^a
D_a\Phi\right] + \int_{\partial B}\chi \tilde{t}^{ab} e_a n_b dS
\label{f11tilde}
\end{eqnarray}
where $F = F^a e_a$ and $\Phi = - A_a \xi^a$ with $A_a$ is the vector
potential of the background black hole.

Finally, for the case where the thermal atmosphere surrounds the box,
$\tilde{T}^{ab}$ outside of the box will have a perfect fluid form, so
eq.(\ref{f11tilde}) further simplifies to
\begin{eqnarray}
\chi_0 F = \int_B \left[\tilde{\rho} e^a D_a \chi + \tilde{q} e^a
D_a\Phi\right] + \int_{\partial B}\chi \tilde{P} e_a n^a dS
\label{f11tilde2}
\end{eqnarray}
For the case of an empty box ($\tilde{\rho} = \tilde{q} = 0$), we obtain
eq.(\ref{F}) used in our analysis in section 5.


\begin{thebibliography}{3}

\bibitem{tbh} R.M. Wald, ``The Thermodynamics of Black Holes'', Living
Reviews in General Relativity (in press); gr-qc//9912119.
\bibitem{bch} J.M. Bardeen, B. Carter, and S.W. Hawking,
Commun. Math. Phys. {\bf 31}, 161 (1973).
\bibitem{iyer1} V. Iyer and  R.M. Wald, Phys. Rev. {\bf D50}, 846 (1994).
\bibitem{waldbook} R.M. Wald, {\em Quantum Field Theory in Curved
Spacetime and Black Hole Thermodynamics}, University of Chicago Press
(Chicago, 1994).
\bibitem{bek} J.D. Bekenstein, Phys. Rev. {\bf D7}, 2333 (1973);
Phys. Rev. {\bf D9}, 3292(1974).
\bibitem{bek2} J.D. Bekenstein, Phys. Rev. {\bf D23}, 287 (1981).
\bibitem{unruhwald} W.G. Unruh and R.M. Wald, Phys. Rev. {\bf D25}, 942
(1982).
\bibitem{bek3} J.D. Bekenstein, Phys. Rev. {\bf D27}, 2262 (1983).
\bibitem{bek4} J.D. Bekenstein, Phys. Rev. {\bf D49}, 1912 (1994).
\bibitem{bek5} J.D. Bekenstein, Phys. Rev. {\bf D60}, 124010 (1999);
gr-qc/9906058.
\bibitem{uw2} W.G. Unruh and R.M. Wald, Phys. Rev. {\bf D27}, 2271
(1983).
\bibitem{mark} M. Pelath and Wald, Phys. Rev. {\bf D60}, 104009 (1999).
\bibitem{bm} J.D. Bekenstein and A.E. Mayo, Phys. Rev. {\bf D61},
024022 (2000).
\bibitem{hod1} S. Hod, Phys. Rev. {\bf D61}, 024018 (2000).
\bibitem{hod2} S. Hod, Phys. Rev. {\bf D61}, 024023 (2000).
\bibitem{muko} T. Shimomura and S. Mukohyama,  Phys. Rev. {\bf D61},
064020 (2000). 
\bibitem{iyer2} V. Iyer and R.M. Wald, Phys. Rev. {\bf D52}, 4430 (1995).
\bibitem{zoupas} R.M. Wald and A. Zoupas, Phys. Rev. {\bf D61}, 084027 (2000).
\bibitem{sud} D. Sudarsky and R.M. Wald, Phys. Rev. {\bf D46}, 1453 (1992).
\bibitem{car} B. Carter, ``Black Hole Equilibrium States'' in {\em
Black Holes}, ed.by B. DeWitt and C. DeWitt, Gordon and Breach (New
York, 1973).
\bibitem{gr} R.M. Wald, {\em  General Relativity} (University of Chicago
Press, Chicago, 1984)
\bibitem{gp} G.W. Gibbons and M.J. Perry, Phys. Rev. Lett. {\bf 36},
985 (1976).
\bibitem{kw} B.S. Kay and R.M. Wald, Phys. Rep. {\bf 207}, 49 (1991).
\bibitem{kay} B.S. Kay, Commun. Math. Phys. {\bf 62}, 55 (1978).
\bibitem{wa} R.M. Wald, Class. Quant. Grav. {\bf 16}, A177 (1999).
\bibitem{ger} R. Geroch, J. Math. Phys. {\bf 12}, 918 (1971).
\end{thebibliography}
\end{document}